\documentclass[letterpaper]{sig-alternate}
\usepackage[top=1in, bottom=1in, left=1in, right=1in]{geometry}

\usepackage{hyperref}
\usepackage{indentfirst}

\usepackage[latin1]{inputenc}

\usepackage{microtype}
\usepackage{graphicx}
\usepackage{amsfonts}
\usepackage{appendix}
\usepackage{url}
\usepackage{multirow}
\usepackage{subfigure}
\usepackage{verbatim}
\usepackage{listings}
\usepackage{balance}
\usepackage{color}
\usepackage{cuted}
\usepackage{color}
\usepackage{pgf}

\begin{document}

\title{How Unique and Traceable are Usernames?}
\author{
\alignauthor
Daniele Perito, Claude Castelluccia, Mohamed Ali Kaafar, Pere Manils\\
  \affaddr{INRIA Rh\^one-Alpes}\\
  \email{\{perito,ccastel,kaafar,manils\}@inrialpes.fr}\\
\alignauthor
}

\maketitle

\begin{abstract}

Suppose you find the same username on different online services, what is the
probability that these usernames refer to the same physical person?  This work
addresses what appears to be a fairly simple question, which has many
implications for anonymity and privacy on the Internet.  One possible way of
estimating this probability would be to look at the public information
associated to the two accounts and try to match them. However, for most
services, these information are chosen by the users themselves and are often
very heterogeneous, possibly false and difficult to collect. Furthermore,
several websites do not disclose any additional public information about users
apart from their usernames (e.g., discussion forums or Blog comments),
nonetheless, they might contain sensitive information about users.

This paper explores the possibility of linking users profiles only by looking
at their usernames.  The intuition is that the probability that two usernames
refer to the same physical person strongly depends on the ``entropy'' of the
username string itself.  Our experiments, based on crawls of real web services,
show that a significant portion of the users' profiles can be linked using
their usernames.  To the best of our knowledge, this is the first time that
usernames are considered as a source of information when profiling users on the
Internet.


\end{abstract}


\section{Introduction}

Online profiling is a serious threat to users privacy. In particular, the
ability to trace users by linking multiple identities from different public
profiles may be of great interest to profilers, advertisers and the like. Indeed,
it might be possible to gather information from different online services and combine it
to sharpen the knowledge of users identities. This knowledge may then be exploited
to perform efficient social phishing or targeted spam, and might be as well
used by advertisers or future employers seeking information.  As it has been
colloquially put by a judge of the US Supreme Court in a recent case about
warrantless GPS
tracking\footnote{http://www.eff.org/press/archives/2010/08/06-0}: ``When it
comes to privacy, the whole may be more revealing than its parts.''

Recent works \cite{www2009,RAID} showed how it is possible to retrieve
users information from different online social networks (OSN). 
All of these works mainly exploit flaws in the OSN's API
design  (e.g., Facebook friend search).
Other approaches \cite{deanonymizingSP09} use the topology of
social network friend graphs to de-anonymize its nodes.

In this paper, we propose a novel methodology that uses 
usernames -an easy to collect information- rather than social graphs to tie user online identities. 
Our technique only assumes knowledge of usernames and it is widely applicable to all web
services that publicly expose usernames.
Our purpose is to show that users' pseudonyms allow simple, yet efficient tracking of online activities.

Recent scraping services' activities illustrate well the threats introduced by
the ability to match up user's pseudonyms on different social
networks~\cite{link_to_nielsen}. For instance, PeekYou.com has lately
applied for a patent for a way to match people's real names to
pseudonyms they use on blogs, OSN services and online forums
\cite{Agrgregatorpatent}. The methodology relies on
public information collected for an user, that might help in matching different
online identities. The algorithm empirically assigns weights to each of the
collected information so as to deem different identities to be the same.
However, the algorithm is ad-hoc and not robust to false or mismatching information.
In light of these recent developments, it is desirable that the research community
investigates the capabilities and limits of these profiling techniques.
This will, in turn, allow for the design of appropriate countermeasures to
protect users' privacy.

In general, profiling unique identities from multiple public profiles is a
challenging task, as information from public profiles is often incorrect,
misleading or altogether missing \cite{footprint}. Techniques designed for the
purpose of profiling need to be robust to these occurrences.

\paragraph{Contributions} 

The contributions of this paper are manifold.
First, we introduce the problem of linking multiple online identities relying only
on usernames.
Second, we devise an analytical
model to estimate the {\em uniqueness} of a username,
which can in turn be used to assign a 
probability that a single username, from two
different online services, refers to the same user. Based on language models
and Markov Chain techniques, our model validates an intuitive observation:
usernames with low ``entropy'' (or to be precise {\em Information Surprisal}) will have higher probabilities of being picked
by multiple persons, whereas higher entropy usernames will be very
unlikely picked by multiple users and refer in the vast majority of the cases to unique users.

Third, we extend this model to cases when usernames are {\em different} across many
online services. In essence, given two usernames our technique returns
the probability that these usernames refer to the same user, and allows
then to effectively trace users identities across multiple
web services using their usernames only. 
While we acknowledge that our technique cannot trace users that choose
unrelated usernames on purpose, experimental data shows that
users tend to choose closely related usernames on different services.
Finally, by applying our technique to subsets of usernames we extracted from real
cases scenarios, we validate and discuss our technique in the wild.

We envision several possible uses of these techniques, not all of them malicious.
In particular, users might use our tool to test how unique their username is and, therefore,
take appropriate decision in case they wish to stay anonymous.
To this extent we provide an online tool that can help users choose appropriate
usernames by measuring how unique and traceable their usernames are. 
The tool is available at \url{http://planete.inrialpes.fr/projects/how-unique-are-your-usernames}.

%

\paragraph{Paper organization}

In Section \ref{overview}, we overview the related work on privacy and
introduce the machine learning tools used in our analysis.  In Section
\ref{unique}, we introduce our measure to estimate the uniqueness of usernames
and in Section \ref{couple}, we extend our model to compute the probability
that two usernames refer to the same person and validate it using the dataset
we collected from eBay and Google (Section \ref{dataset}).  Different
techniques are introduced and evaluated. Finally, in Section
\ref{sec:discussion} we discuss potential impact of our proposed techniques and
present some possible countermeasures.

%
%
%

\section{Related work and Background}
\label{overview}

\subsection{Related Work}

\paragraph{Tracking OSNs users}

In \cite{footprint} the authors propose to use what they call the {\em online
social footprint} to profile users on the Internet. This footprint would be the
collection of all little pieces of information that each user leaves on
web services and OSNs. While the idea is promising this appears to be only a preliminary work and
no model, implementation or validation is given.

Similarly in \cite{www2009}, Bilge et al. discuss how to link the
membership of users to two different online social networks. Noticing that
there might be discrepancies in the information provided by a single user in
two social networks, the authors rely on Google search results to decide the
equivalence of selected fields of interest (as for assigning uniqueness of a
user). Typically, the input of their algorithm is the name and surname of a
user, that is augmented by the education/occupation as provided in two
different social networks. They use such input to start two separate Google
searches, and if both appear in the first top three hits, these are deemed to
be equivalent. The corresponding users are consequently identified as a single
user on both social networking sites.  Bilge et al.'s work illustrates well how
challenging the process of identifying users from multiple public profiles is.
Despite the usage of customized crawler and parser for each social network, the
heterogeneity of information as provided by users (if correct) makes the
process hard to deploy, if not unfeasible, at a large scale.


\paragraph{Record linkage}

Record linkage (RL)(or alternatively Entity Resolution) \cite{rl, rl_comparison} refers to the task of finding records that refer to the
same entity in two or more databases. This is a common task when databases of users
records are merged. For example, after two companies merge they might also want to
merge their databases and find duplicate entries.
Record linkage is needed in this case if there are no unique identifiers
available (e.g., social security numbers).

In RL terminology two records that have been matched are said to be {\em
linked}. 
For the purpose of linking profiles using usernames, we test several
RL techniques and compare their performance to the ones introduced in this paper.
However, differently from the record linkage problem, in our setup a
complete match of two different usernames does not necessarily indicate a
positive identification. Furthermore, the application of record linkage
techniques to link public online user profiles is novel to the best of our
knowledge and presents several challenges of its own.

\paragraph{Tracking browsers across sessions}

Another related problem is the fingerprinting of web clients. Usually, ad servers set unique cookies on the browsers to allow for easy
tracking of users between HTTP requests.  A simple and straightforward practise
on browsers to limit the risk of re-identification is to restrict or disable
the use of third-party cookies. However, recent research \cite{EFF} has shown
that different browser installations might contain enough unique features or
``entropy'' to allow for re-identification even in the absence of long lived
unique identifiers like cookies.

\paragraph{De-anonymizing sparse database and graph data}

\cite{deanonymizingSP09} proposes an identification algorithm targeting
anonymized social network graphs. The main idea of this work is to de-anonymize
online social graph based on information acquired from a secondary social
network users are known to belong to as well. Similarity identified in the
network topologies of both services allows then to identify users belonging to
the anonymized graph.  
 
\subsection{Background}

\subsubsection{Information Surprisal}

Self-information or Information Surprisal measures the amount of information
associated to a specific outcome of a random variable. If $X$ is a random variable
and $x$ one possible outcome, we denote the information surprisal of $x$ as $I(x)$.
Information Surprisal is computed as $I(x) = -\log_2(P(x))$ and hence depends
only on the probability of $x$. The smaller the probability of $x$ the higher
is the associated surprisal.
Entropy, on the other
hand, measures the information associated to a random variable (regardless of any
specific outcome), denoted $H(X)$. Entropy and Surprisal are deeply related
as entropy can be seen as the expected value of the information surprisal, $H(X)=E(I(X))$.
Both are usually measured in bits.

Suppose there exists a discrete random variable that models the distribution of
usernames in a population, call this variable $U$.
The random variable $U$ follows a probability mass function $P_U$
that associates to each username $u$ a probability $P(u)$. In this context,
the information surprisal of $P(u)$ is the
amount of identifying information associated to a username $u$. Every bit
of surprisal adds one bit of identifying information and thus allows to cut the
population in which $u$ might lie in half. 

If we assume that there are $W$ users in a population, then a username $u$
identifies {\em uniquely} a user in the population if $I(P(u)) > \log_2(W)$.  In this sense,
information surprisal gives a measure of the ``uniqueness'' of a username $u$
and it is the measure we are going to use in this work.  The challenge lies in
estimating the probability $P(u)$, which we will address in Section
\ref{unique}.

Our treatment of information surprisal and
its association to privacy is similar to the one recently suggested in
\cite{EFF} in the context of fingerprinting browsers.

%

\section{The Dataset}
\label{dataset}

Our study was conducted on several different lists of usernames: (a) a list of
$3.5$ million usernames gathered from public Google profiles; (b) a list of
$6.5$ million usernames gathered from eBay accounts; (c) a list of 16000
thousand username gathered from our research center LDAP directory; (d) two
large username lists found online used in a previous study from Dell'Amico et
al.~\cite{infocom10}: a ``finnish'' dataset and a list of usernames collected
from Myspace.  

The ``finnish'' dataset comes from a list publicly disclosed in October
2007\footnote{\url{http://www.f-secure.com/weblog/archives/00001293.html}}. The
dataset contains usernames, email addresses and passwords of almost 79000 user
accounts. This information has been collected from -most likely by hacking- the
servers of several Finnish web forums. The MySpace dataset comes from a
phishing attack, setting a fake MySpace login web page. This data has been
discolsed in October 2006 and 
it contains more than 30000 unique usernames.

The use we made of these datasets was threefold.  First, we used the combined
list of $10$ million usernames (from eBay and Google) to train our Markov Chain
model needed for the probability estimations.  Second, we used the information
on Google profiles to gather ground truth evidence and test our technique to
link multiple public profiles even in case of slightly different usernames (Section \ref{couple}).
Third, we used all the datasets to characterize username uniqueness and depict
Surprisal information distributions as seen in the wild. Our objective here is
to validate our techniques on several datasets, where users come from widely
distributed locations and may have different habits as for web services usage
and usernames' choices. 

Notably, a feature of Google Profiles \footnote{\url{http://www.google.com/profiles}}, allowed us
to build a {\em ground truth} we used for validation purposes.
In fact, users on Google Profiles can optionally decide to provide a list of their
other accounts on different OSNs and web services.
This provided us with a ground truth, for a subset of all profiles,
of linked accounts and usernames.

In our experiments we observed that web services differ significantly in their
username policies.  However, almost all services share a common alphabet of
letters and numbers.  Analyzing our most complete set of $10$ million distinct
usernames it appears clear that $85\%$ of the users choose alphanumerical {\em
only} usernames that thus comply to all username policy.
This fact is of interest when evaluating the applicability of the techniques
explained in this work.

\section{Estimating Username Uniqueness}
\label{unique}

As we explained above, we would like to have a measure of username {\em uniqueness},
which can quantify the amount of identifying information each username carries.
Information Surprisal is a measure, expressed in bits, that serves this 
purpose.
However, in order to compute the Information Surprisal associated to usernames, we need
a way to estimate the probability $P(u)$ for each username $u$.

A naive way to estimate $P(u)$, given a dataset of
usernames coming from different services, would be to use
Maximum Likelihood Estimation (MLE). If we have $N$ usernames then we can
estimate the probability of each username $u$ as $\frac{count(u)}{N}$, if $u$
belongs to our dataset, and $0$ otherwise.  Where $count(u)$ is simply the
number of occurrences of $u$ in the sample. In this case we are assigning
maximum probability to the observed samples and zero to all the others.
This approach has several drawback, but the most severe is that
it cannot be used to give any estimation for the usernames not in the sample.
Furthermore, the estimation given is very coarse.

Instead, we would like to have a probability estimation that allows us to give
estimate probabilities for usernames we have never encountered. 
 Markov-Chains
have been successfully used to extrapolate knowledge of human language from small
corpuses of text. In our case, we apply Markov Chain techniques on usernames
to estimate their probability.

\subsection{Estimating username probabilities with \\ Markov Chains}


Markov models are successfully used in many machine learning techniques that
need to predict human generated sequences of words, as in speech recognition
\cite{stat_nlp}. In a very common machine learning problem, one is faced with
the challenge of predicting the next word in a sentence. If for example the
sentence is {\em ``The quick brown fox''}, the word {\em jumps} would be a more
likely candidate than {\em car}. This problem is usually referred to as {\em
Shannon Game} following Shannon's seminal work on the topic\cite{shannon51}.  This
task is usually tackled using Markov-Chains and modeling the probability of the word
{\em jumps} depending of a number of words preceding it.

In our scenario, the same technique can be used to estimate the probability of
username strings instead of sentences.  For example, if one is given the
beginning of a username like \texttt{sara}, it is possible to predict that the
next character in the username will likely be \texttt{h}.  Notably Markov-Chain
techniques have been successfully used to build password crackers \cite{Narayanan05fastdictionary}
and analyse the strength of passwords \cite{infocom10}.

Without loss of generality, the probability of a given string $c_1,...,c_n$ can be written as 
$\Pi_{i=1}^{n}P(c_i|c_1,...,c_{i-1})$.
In order to make calculation possible a Markovian assumption is introduced:
to compute the probability of the next character, only the previous $k$
characters are considered.  This assumption is important because it
simplifies the problem of learning the model from a dataset.  The
probability of any given username can be expressed as:

$$P(c_1,...,c_n) = \Pi_{i=1}^nP(c_i|c_{i-k+1},...,c_{i-1})$$

To utilize Markov-Chain for our task we need to estimate, in a learning phase,
the model parameters (the conditional probabilities) using a suitable dataset.
In our experiments we used the database of approximately $10$ million
usernames populated by collecting Google public profiles and eBay user accounts
(see Section \ref{dataset}).

In general, the conditional probabilities are computed as: 

$$P(c_i|c_{i-k+1},...,c_{i-1}) =
\frac{count(c_{i-k+1},...,c_{i-1},c_i)}{count(c_{i-k+1},...,c_{i-1})}$$ by
counting the number of $n$-grams that contain character $c_i$ and dividing it
by the total number of $n-1$-grams without the character $c_i$.
Where an $n$-gram is simply a sequence of $n$ characters.

Markov-Chain techniques benefit from the use of longer $n$-grams, because
longer ``histories'' can be captured. However longer $n$-grams result into 
an exponential decrease of the number of samples for each $n$-gram.
In our experiments we used $5$-grams for the computation of conditional
probabilities. 

Once we have calculated $P(u)$, we can trivially compute the information
surprisal of $u$ as $-\log_2(P(u))$.

In Appendix \ref{app:unique} we give a different, yet related, probabilistic explanation
of username uniqueness.

%
 
\begin{figure} [t] \centering
\includegraphics[scale=0.62]{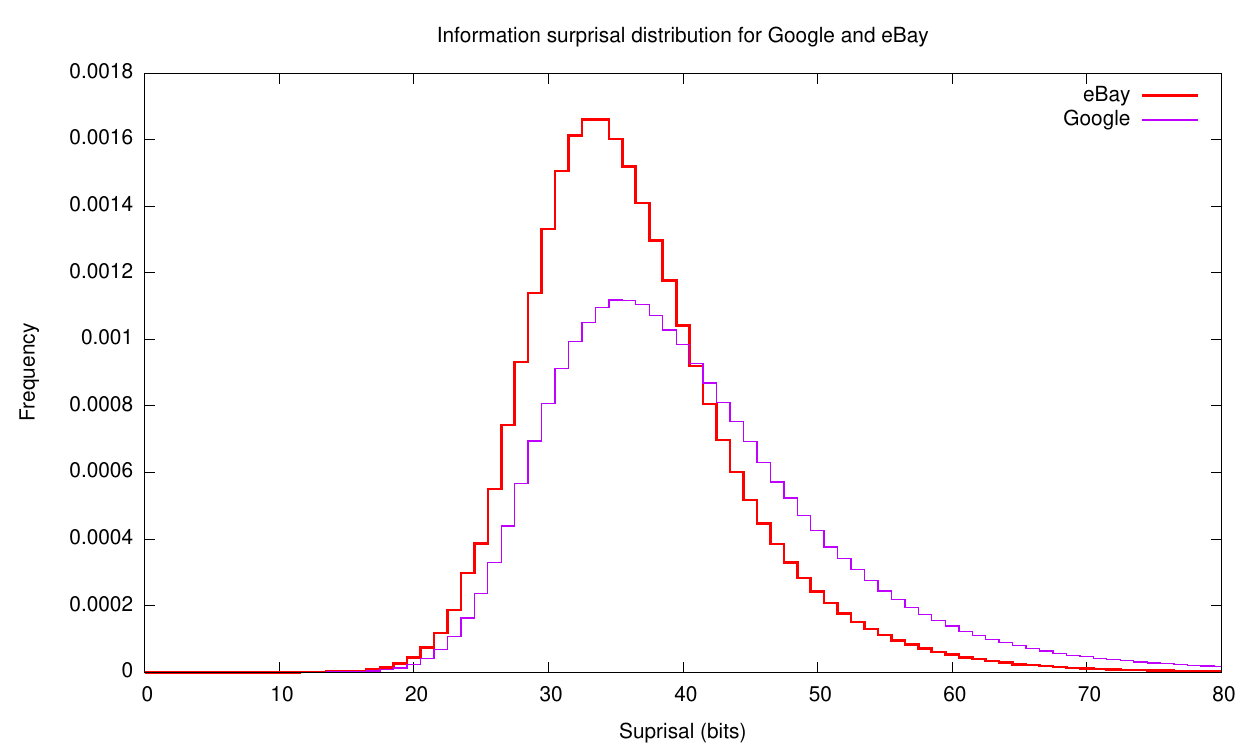}
\caption{Surprisal distribution for eBay and Google usernames}
\label{fig:surprisal}
\end{figure}

\begin{figure} [t] \centering
\includegraphics[scale=0.62]{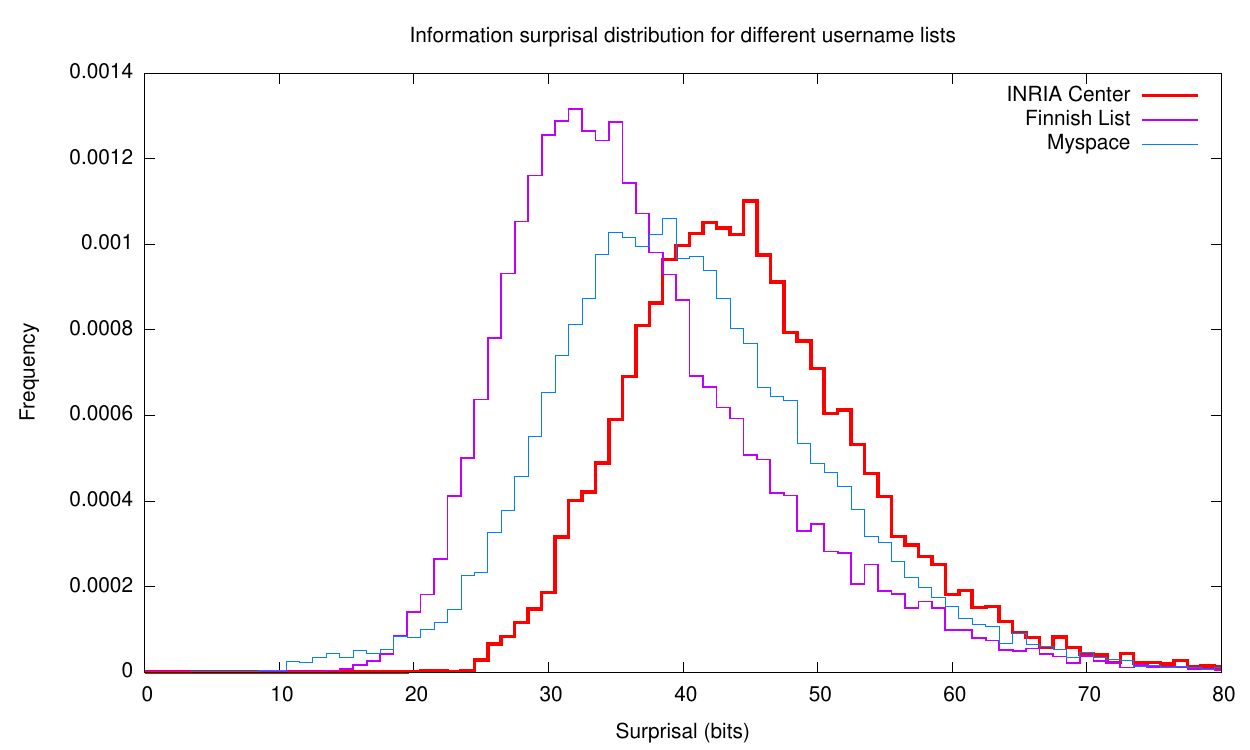}
\caption{Surprisal distribution for other services}
\label{fig:surprisal_inria}
\end{figure}

\begin{figure} [t] \centering
\includegraphics[scale=0.21]{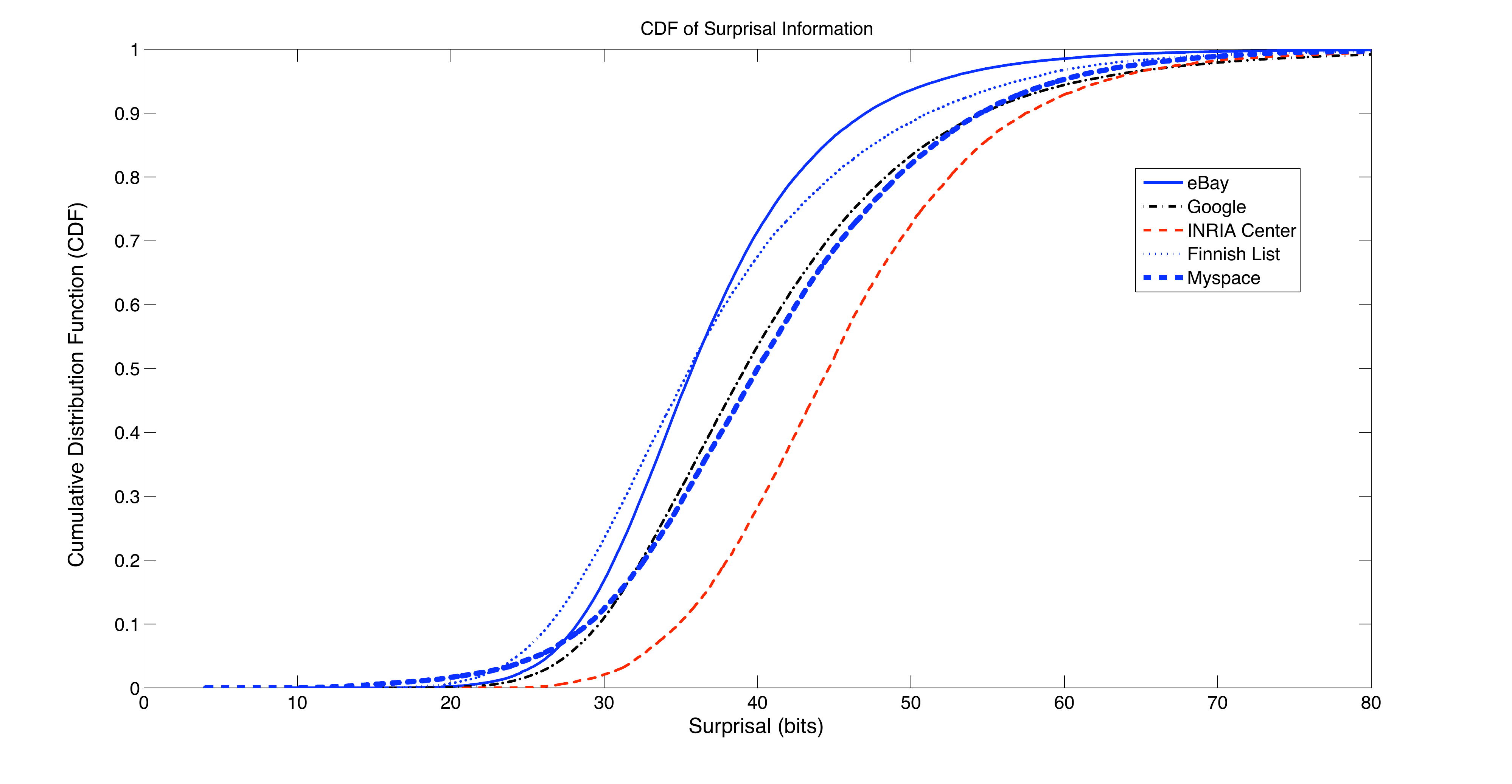} \caption{Cumulative
distribution function for the surprisal of all the services}
\label{fig:cdf-surprisal} \end{figure}

\subsection{Experiments}

We conducted experiments to estimate the surprisal of the usernames in our
dataset and hence how unique and identifying they are.
As explained above, our Markov-Chain model was trained using the combined
$10$ million usernames gathered from eBay and Google.
The dataset was used for both training and testing by using leave-one-out
cross validation. Essentially, when computing the probability of a username
$u$ using our Markov-Chain tool, we {\em excluded} $u$ from the model the
occurrence counts. This way, the probability
estimation for $u$ depended on all the other usernames but $u$. 

We computed information surprisal for all the usernames in our dataset and the results are
shown in Figure \ref{fig:surprisal}.
The entropy of both distributions is higher than $35$ bits which would suggest that,
on average, usernames are extremely unique identifiers.

Notice the overlap in the distributions that might indicate that our surprisal measure
is stable across different services. Notably, the two services have largely different
username creation policies, with eBay accepting usernames as short as $3$ characters from
a wider alphabet and Google giving more restrictions to the users.
Also, the account creation interfaces vary greatly across the two services. In fact, Google
offers a feature that suggests usernames to new users derived from first and last names.
Probably this is the reason why Google usernames have a higher Information Surprisal (see Figure \ref{fig:cdf-surprisal}).
It must also be noted that both services have hundreds of millions
of reported users. This raises the entropy of both distributions:
as the number of users increases they are forced to choose usernames with higher
entropies to find {\em available} ones. 
Overall it appears clear that usernames constitute highly identifying piece of
information, that can be used to track users across websites.

In Figure \ref{fig:surprisal_inria} we plot information surprisal for three
datasets gathered from different services.
This graph is motivated by our need to understand how much surprisal varies across
services.

The results are similar to the ones obtained for eBay and Google usernames. 
The Finnish list is noteworthy, these usernames come from different Finnish
forums and most likely belong to Finnish users. However, Suomi (the official
language in Finland) shares almost no common roots with Roman or Anglo-Saxon
languages. This can be seen as a good representative
of the stability of our estimation for different
languages.

Furthermore, notice that the dataset coming from our own research center
(INRIA) has a higher surprisal than all the other datasets. While there are a
possible number of explanations for this, the most likely one comes from the
username creation policies in place that require usernames to be the
concatenation of first and last name.  The high surprisal comes despite the
fact that the center has only around $16000$ registered usernames and lack
of availability does not pressure users to choose more unique usernames.

Comparing the distributions of Information surprisal of our different datasets
is enlightening, as illustrated in Figure \ref{fig:cdf-surprisal}. This
confirms that usernames collected from the INRIA center exhibit the highest
information surprisal, with almost 75\% of usernames with a surprisal higher
than 40 bits. We also observe that both Google and MySpace CDF curves closely
match. In all cases, it is worth noticing that the maximum (resp. the minimum)
fraction of usernames that do exhibit an information surprisal less than 30
bits is 25\% (resp. less than 5\%). This shows that a vast majority of users
from our datasets can be uniquely identified among a population of 1 billion
users, relying only on their usernames.

\section{Username couples linkage}
\label{couple}

The technique explained above can only estimate the uniqueness of a single
username across multiple web services. However, there are cases in which users,
either willingly or forced by availability, decide to change their username. 

\begin{figure}[ht]
\label{fig:leven}
\centering
\subfigure[Levenshtein distance distribution for linked username couples (set $L$, $|L| = 10^4$)]{
    \includegraphics[scale=0.5]{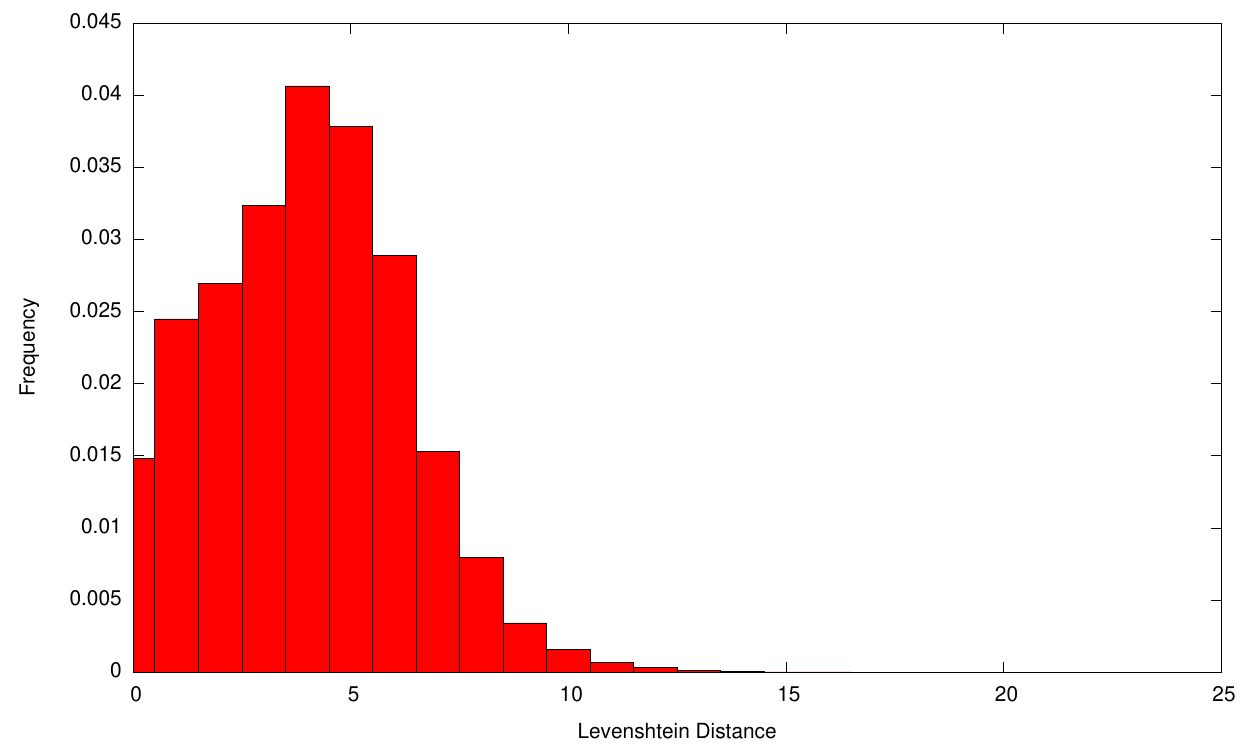}
    \label{fig:same_couple}
}
\subfigure[Levenshtein distance distribution for non-linked username couples (set $NL$, $|NL| = 10^4$)]{
    \includegraphics[scale=0.5]{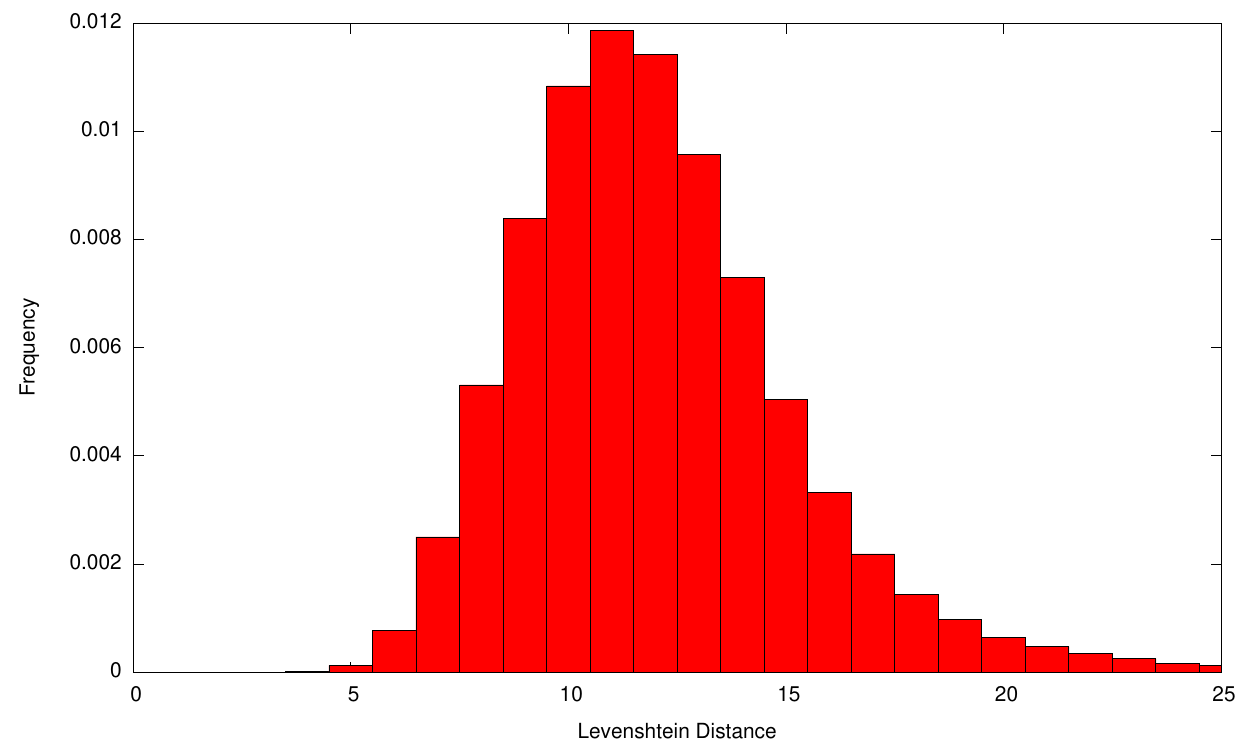}
    \label{fig:impo_couple}
}
\caption[Caption]{Levenshtein distance distribution for username couples gathered from $3$ million Google profiles}
\end{figure}

We would like to know whether users change their usernames in
any predictable and traceable way.
In Figure \ref{fig:same_couple} and \ref{fig:impo_couple} is plotted the distribution of the Levenshtein (or Edit) Distance for
username couples.  In particular, Figure \ref{fig:same_couple} depicts the
distribution for $10^4$ username couples we can verify to belong to single users
(we call this set $L$ for {\em linked}), using our dataset.  On the other hand,
Figure \ref{fig:impo_couple} shows the distribution for a sample of random
username couples that do not belong to a single user (we call this set $NL$ for
{\em non-linked}).  In the first case the mean distance is $4.2$ and the
standard deviation is $2.2$, in the second case the mean Levenshtein distance is $12$ and the
standard deviation is $3.1$. 
Clearly, linked usernames are much closer to each other than non linked ones.
This suggests that, in many occurrences, users choose usernames that are
related to each other.
The difference in the two distributions is remarkable and so it might be
possible to estimate the probability that two different usernames are used by
the same person or, in record linkage terminology, to {\em link} different
usernames. 

However, as illustrated in Section \ref{unique}, and differently from record linkage, an almost perfect username match does not always indicate that the
two usernames belong to the same person.
The probability that two usernames, let's say \texttt{sarah} and \texttt{sarah2}, are linked (we call it $P_{same}(sarah, sarah2)$) should depend
on:

\begin{enumerate} \item how much `information' there is in the common part of the usernames (in this
case \texttt{sarah}) and, 
\item how likely is that a user will change one username into
the other (in this case the addition of a \texttt{2} at the end). 
\end{enumerate}

We will show two different novel approaches at solving this problem. The first approach
uses a combination of Markov Chains and a weighted Levenshtein Distance using probabilities.
The second approach makes use of the theory and techniques used for information retrieval
in order to compute document similarity, specifically using TF-IDF.

We compare these two techniques to record linkage techniques for a base-line comparison.
Specifically we use string-only metrics
like the normalized Levenshtein Distance and Jaro distance to link username couples.

\paragraph{Method 1: Linkage using Markov-Chains and LD}

First of all, we need to compute the probability of a certain
username $u_1$ being changed into $u_2$. We denote this probability as $P(u_2|u_1)$.
Going back to our original example, $P(sarah2|sarah)$ is equal to the
probability of adding the character $2$ at the end of the string
\texttt{sarah}. This same principle can be extended to deletion and
substitution.  In general, if two strings $u_1$ and $u_2$ differ by a sequence
of basic operations $o_1, o_2,...,o_n$, we can estimate $P(u_2| u_1) = P(u_1 | u_2) =
p(o_1)p(o_2)...p(o_n)$.

In order to estimate the probability that username $u_1$ and $u_2$ belong to
the same person, we need to consider that there are two different possibilities
on how $u_1$ and $u_2$ were chosen in the first place.  The first possibility is that they
were picked independently by two different users. The second possibility is
that they were picked by the same user, hence they are not independent.

In the former case we can compute $P(u_1 \wedge u_2)$ as  $P(u_1)*P(u_2)$ 
since we can assume independence. In the latter, $P(u_1 \wedge u_2)$ equals $P(u_1)*P(u_2|u_1)$
in case the user is the same. Note that using Markov Chains and the our estimation of $P(u_2|u_1)$, 
we can compute all the terms involved.
Estimating the probability $P_{same}(u_1, u_2)$ is now a matter of estimating and
comparing the two probabilities above.

%
%

The formula for $P_{same}(u_1, u_2)$ is derived from the
probability $P(u_1 \wedge u_2)$ using Bayes' Theorem.
In fact, we can rewrite the probability above as $P(u_1 \wedge u_2 | S)$ where the random
variable $S$ can have values $0$ or $1$ and it is $1$ if $u_1$ and $u_2$ belong to the same person
and $0$ otherwise.
Hence without loss of generality:
$$P(S|u_1 \wedge u_2) = \frac{P(u_1 \wedge u_2 | S)P(S)}{\sum_{S={0,1}}(P(u_1 \wedge u_2 | S)*P(S))}$$
which leads to $P(S=1|u_1 \wedge u_2)$ equal to
$$ \frac{P(u_1)P(u_2|u_1)P(S=1)}{P(u_1)P(u_2)P(S=0)+P(u_1)P(u_2|u_1)P(S=1)}$$
where $P(S=1)$ is the probability of two usernames belonging to the same
person, regardless of the usernames.  We can estimate this probability to be
$\frac{1}{W}$, where $W$ is the population size. Conversely $P(S=0) = \frac{W-1}{W}$.
And so we can rewrite $P_{same}(u_1, u_2)$ as $P(S=1|u_1 \wedge u_2)$ equal to
$$\frac{P(u_1)P(u_2|u_1)}{W*P(u_1)P(u_2)\frac{W-1}{W}+W*P(u_1)P(u_2|u_1)\frac{1}{W}}$$
Please note that when $u_1 = u_2 = u$ then the formula above becomes $$P_{same}(u, u) = \frac{1}{(W-1)P(u)+1} = P_{uniq}(u)$$ which
is exactly the same estimation we devised for the username uniqueness in Appendix.


\paragraph{Method 2: Linkage using TF-IDF}

In this case we use a well known information retrieval tool called TF-IDF.
However, TF-IDF similarity measures the distance between two documents (or a
search query and a document), which are set of words. 

The term frequency-inverse document frequency (TF-IDF) is a weight used to evaluate
how important is a word to a document that belongs to a corpus \cite{TFIDF}. 
The weight assigned to a word increases proportionally to the number of times
the word appears in the corpus but the importance decreases for common words in
the corpus.

If we have a collection of documents $D$ in which each document $d \in D$ is a
set of terms, then we can compute the {\em term frequency} of term $t_i \in d$
as: $tf_{i,j} = \frac{n_{i,j}}{\Sigma_k n_{k,j}}$ where $n_{i,j}$ is the number
of times term $t_i$ appears in document $d_j$.  The {\em inverse document
frequency} of a term $t_i$ in a corpus $D$ is $idf_i = \frac{|D|}{c_i}$ where
$c_i$ is the number of documents in the corpus that contain the term $t_i$.
The TF-IDF is computed as $(tf-idf)_{i,j} = tf_{i,j}idf_i$.
The TF-IDF is often used to measure the similarity between two documents, say
$d$ and $d'$, in the following way: first the TF-IDF is computed over all the
term in $d$ and $d'$ and the results are stored in two vectors $v$ and $v'$;
then the similarity between the two vectors is computed, for example using a cosine
similarity measure $sim(d,d') = \frac{v \cdot v'}{\|v\|\|v'\|}$.

In our case we need to
measure the distance between usernames composed of a single string.
The way we solved this problem is pragmatical: we consider all possible
substrings, of size $q$, of a string $u$ to be a document $d_u$. Where $d_u$ can be seen as the {\em
building blocks} of the string $u$.  The similarity between username $u_1$ and
$u_2$ is computed using the similarity measure described above. This
similarity measure is referred to in the literature as $q$-gram similarity
\cite{qgrams}, however it has been proposed for fuzzy string matching in
database applications and its application to online profiling is novel.

\paragraph{Method 3: String Only Similarity Metrics from Record Linkage}
\label{back:tfidf}

The Levenshtein (or edit) distance (LD) measures the similarity between two
strings of different or equal length. It is defined as the minimum number of
basic operations (deletion, insertion and substitution) needed to edit one string
into another.  The Levenshtein distance is a useful tool
but its interpretation is not always clear in practice.  For example, consider
the case of the usernames \texttt{alice} and \texttt{malice}, in comparison to
the couple \texttt{vonneumann} and \texttt{jvonneumann}. Both couples have a LD
of $1$ but in the latter case the two usernames are clearly more related than
in the former. To cope with these cases a normalized Levenshtein distance (NLD)
is used instead. While there are different methods used to normalize the LD
between two strings, in our experimentations we use the following formula: $NLD
= 1 - \frac{LD}{max(len(u_1), len(u_2))}$. Note that a NLD is always a number
between $0$ and $1$ since the LD can be at most equal to the length of the
longest string. Note also that the longer $u_1$ or $u_2$ are the closer
$NLD$ approaches one.

The Jaro distance \cite{jaro} is yet another measure of similarity between two
strings and it is mainly used in the area of record linkage. The distance is
normalized and goes from $0$ to $1$ with $1$ indicating an exact match.
We will use it as a base-line comparison with our novel approaches.
However, because of lack of space, we will not explain it in detail.

\subsection{Validation}

Our goal is to assess how accurately usernames can be used to link two different accounts.
For this purpose we design and build a classifier to separate the two
sets $L$ and $NL$, respectively of linked usernames and non-linked usernames.

For our tests the ground-truth evidence was gathered from Google Profiles and the
size the number of linked username couples $|L|$ is 10000.
In order to fairly estimate the performance of the classifier in a real world
scenario we also randomly paired 10000 non-linked usernames to generate the $NL$ set.

The username couples were separated, shuffled and a list of
usernames derived from $L$ and $NL$ was constructed.  The task of the
classifier is to re-link the usernames in $L$ maximizing the username
couples correctly linked while linking as few incorrect couples as possible. 

In practise for each username in the list our program computed the
distance to any other username and kept only the link to the {\em single}
username with highest similarity.
If this value is above a threshold then the candidate couple is considered
{\em linked} otherwise {\em non-linked}.

\subsubsection{Measuring the performance of our binary classifier}

Binary classifiers are primarily evaluated in terms of {\em Precision} and {\em
Recall}, where precision is defined in terms of true positives ($TP$) and false
positives ($FP$) as follows $precision = \frac{TP}{TP+FP}$ and recall takes
into account the true positives compared to false negatives $recall =
\frac{TP}{TP+FN}$.  The recall is the proportion of usernames that
where correctly classified as unique ($TP$) out of all unique usernames
($TP+FN$).  In addition to those two measures, we will also use Accuracy
defined, with the addition of true negatives ($TN$) as $accuracy = \frac{TP+TN}{TP+TN+FP+FN}$.

In our case, we are interested in finding usernames couples that are actually linked
(true positives) while minimizing the number of couples that are linked by
mistake (false positives).  Precision for us is a measure of exactness or
fidelity and higher precision means less profiles linked by mistake.  Recall
measures how complete our tool is, which is the ratio of linked profiles
that are found out of all linked ones.
Precision and recall are usually shown together in a precision/recall graph.
The reason is that they are often closely related: a classifier with high
recall usually has sub-optimal precision while one with high precision has lower
recall. An ideal classifier has both a high precision and recall of $1$.

\begin{figure} [ht!] \centering
\includegraphics[scale=0.28, angle=-90]{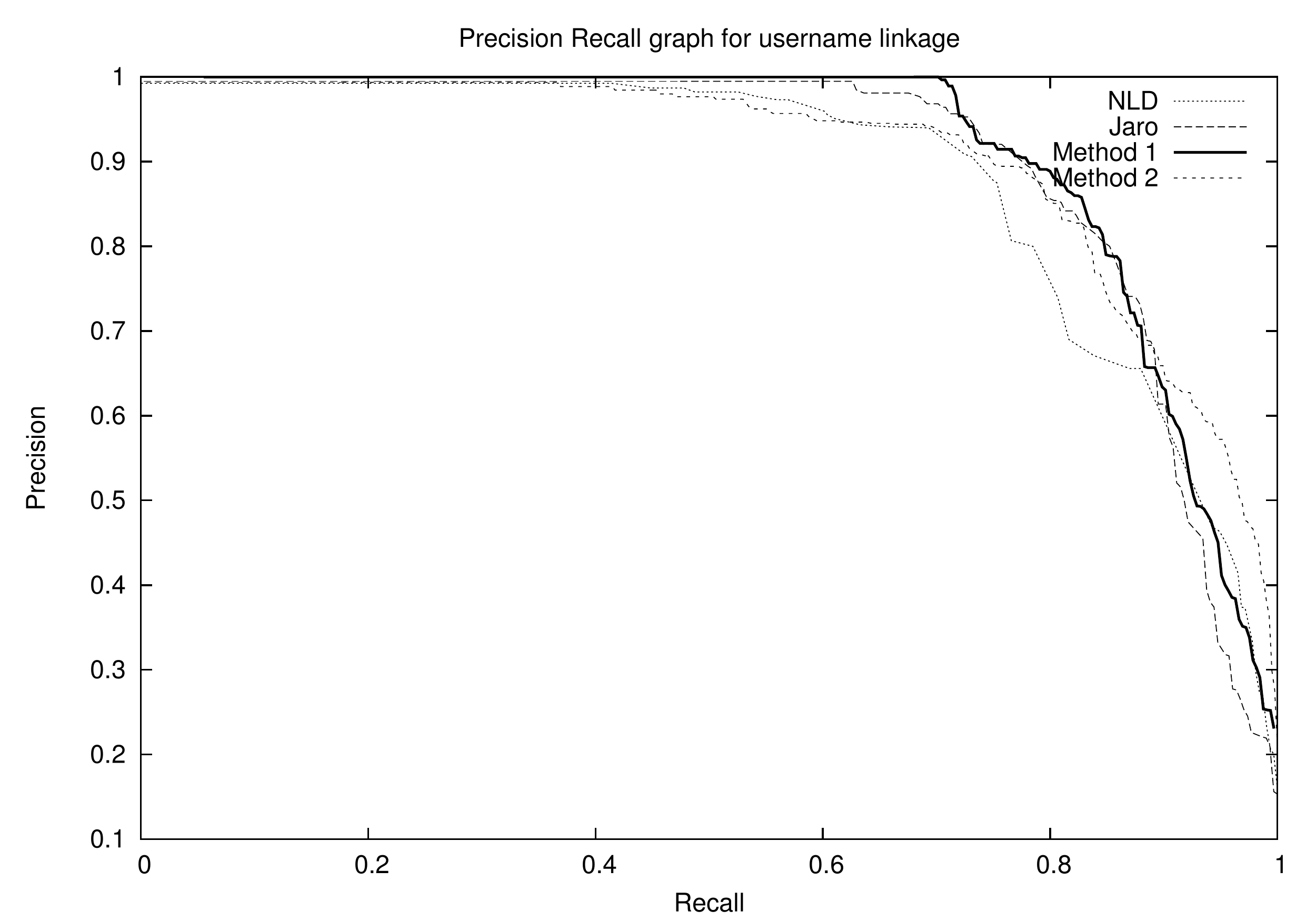}
\caption{Precision and recall for username Linkage}
\label{fig:pr_couples}
\end{figure}

Our classifier looks for potentially matching usernames.
Once a set of potential matches is identified our scoring algorithms are used
to calculate how likely it is that the two usernames represent the same real identity.
By using
our labeled test data, score thresholds can be selected that yield a desired
trade-off between recall and precision. 
Figure \ref{fig:pr_couples} shows the precision and recall of the two methods
discussed in this paper and known string metrics (Jaro and NLD) at various threshold levels.

In general the metric based on Markov models outperforms the other metrics.
Our Markov-Chain method has the advantage of having the highest precision values
especially at recalls up $0.71$. Remember that a recall of $0.71$ means that $71\%$ of all matching
username couples have been successfully linked.
Depending on the application, one might favor TF-IDF based approach (method 2) which has good precision
at higher recalls or the Markov chain approach (method 1) which has the highest precision up to recall $0.7$.

\begin{figure} [tb] \centering
\includegraphics[scale=0.28]{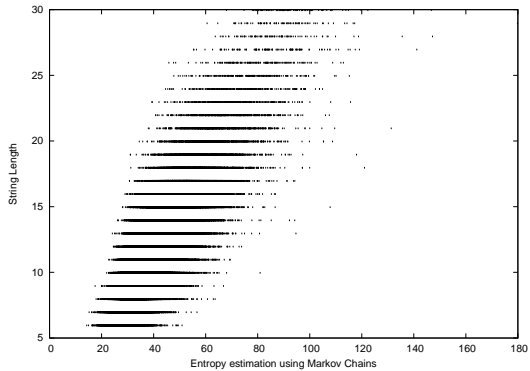}
\caption{Scatter plot for the entropy estimation using Markov Chains and string length}
\label{fig:scatter}
\end{figure}

The string metrics (NLD and Jaro) perform surprisingly well in the task of matching different
usernames. This is probably because, as shown in Figure 4, non-linked
usernames tend to have higher mean distances between themselves than linked
usernames.
Both of these string-only-metric tools assign a positive weight for close strings
and normalize it according to the maximum length of the strings.
Hence, one possible explanation of the performance of NLD and Jaro distances
is that the string length models sufficiently well the surprisal of a string
for the purpose of username linkage.
Indeed, Figure \ref{fig:scatter} shows a scatter plot of the entropy as computed
by our uniqueness metric in comparison to the length of the strings.
The graph clearly shows a central area of correlation between the two metrics and
this is reflected by a high Pearson correlation between the two samples of $0.801$.

%

\subsubsection{Discussion of Results}

Our results show that it is possible, with high precision, to link
accounts solely based on usernames. This is due to the high average entropy
of usernames and the fact that users tend to choose usernames that are related
to each other.
Clearly users could completely change their username for each service they
use and, in this case, our technique would be rendered useless.
However, our analysis shows that users indeed choose similar and
high entropy usernames. 
This phenomenon can be seen as related to the much more
studied password reuse phenomenon \cite{1242661} that plagues web services.
Users tend to reuse a small subset of passwords on $3.9$ services on average, which can be explained
by the difficulty of remembering multiple passwords. 
The same might hold true for usernames, with the notable exception that
usernames can be freely written and do not have to be securely stored.

This technique might be used by profilers and advertiser trying to
link multiple online identities together in order to sharpen their knowledge
of the users. By crawling multiple web services and OSNs (a crawl of 100M Facebook
profiles has already been made available on BitTorrent) profilers could obtain lists of 
accounts with their associated usernames. These usernames could be then used to
link the accounts using the techniques underlined in the previous section.

\subsubsection{Addressing Possible Limitations}

The linked username couples we used as ground truth have been gathered from Google Profiles.
We have shown how that, in this sample, the users tend to choose related usernames.
However, one might argue that this sample might not be sufficiently representative of
the whole population. Indeed Google Profiles users might be least 
concerned about privacy and show a preference of being traceable by
posting their information on their Profiles.

We were not able to test our tool in linking profiles to certain types of web
services in which users are more privacy aware, like dating and medical
websites (e.g., WebMD).  This was due to the difficulty of gathering ground
truth evidence for this class of services.  However, even if we assume that
users choose completely unrelated usernames for different websites, our tool might
still be used.  In fact, it might be the case that a user is registered on
multiple dating websites with similar usernames.  Those profiles might be
linked together with our tool and more complete information about the user
might be found. For example, a date of birth on a website might be linked
with a city of residence and a first name on another, leading to real world
identification.
We acknowledge that, without evidence, this is only speculation and
a more thorough analysis is left for future work.

\subsubsection{Possible Improvements}

Finding linked usernames in a population requires time that is quadratic in the
population size, as all possible couples must be tested for similarity.
This might be too costly if one has millions of usernames to
match. 
A solution to this problem is to divide the matching task in two phases. First, divide
usernames in clusters that are likely be linked.  For example, one could choose
usernames that share at least one $n$-gram, thus restricting the number of
combinations that need to be tried.  Second, test all possible
combinations within a cluster.  


Another possible improvement is to use a hybrid approach in which different
similarity metrics are combined to obtain a single similarity
\cite{rl_comparison}. For instance one could use different similarity metrics (TF-IDF, Markov, Jaro, etc.)
to compose a feature vector that can be then classified using machine learning
techniques like SVMs \cite{svm}. Such hybrid approaches are known to perform
better in the record linkage tasks \cite{rl_comparison}.
However, we did not test or implement such approaches and their application
to linking online identities is left as future work.

\section{Relevant username statistics}
\label{stats}

This section contains username statistics that complement the experiments
we proposed and justify our technique in more practical scenarios.


\paragraph {How do people choose their username?}

We now aim to exploit our Google profiles dataset to verify
whether people use their real name to compose pseudonyms as usernames.  If this
the case an attacker might try to generate likely usernames for a victim and
track the victim on multiple web services using the techniques explained above
to determine username uniqueness and linkage.  Our analysis is then based on
first and last names as provided by users in their profiles. We discard from
the original dataset names provided with strings containing non Latin
characters. These names cannot be mapped to a username according to the Google
policy and so we restrict our study to the Latin alphabet (\emph{a-z}).  For
simplicity, we also considered names composed by two words (i.e. both first
and last names are provided). After this filtering, we ended up with $2.6$M couples of
names and usernames.
%





We decomposed the name into two words that we refer
to as $w_1$ and $w_2$. According to Google profiles policy, $w_1$ (resp. $w_2$)
refers to the first name (resp. last name).  We first performed a preliminary
matching using Perl's regular expressions to check whether usernames contain a
combination of $w_1$, $w_2$ and digits. Results are shown in
Table~\ref{tab:nbu_partial_match}.

\begin{table} [ht!]
  \centering
      \scriptsize
      \begin{tabular}{|c|c|c|}
        \hline
        Matching Condition & \# Usernames & Percentage (\%) \\
        \hline
        \hline
         $w_1$ and $w_2$ and $d$ &  207K  & 7.93 \\
        \hline
         $w_1$ and $w_2$ &  774K &  29.63 \\
        \hline
         $w_1$ and $d$ & 142K  & 5.47\\
        \hline
         $w_2$ and $d$ & 132K &  5.06\\
        \hline
         $w_1$ &  241K &  9.24\\
        \hline
         $w_2$ &  323K &  12.38\\
        \hline
         Not matching & 792K &  30.3\\
        \hline 
        \hline
         Total & 2,6M & 100\\
        \hline
      \end{tabular}
  \caption{Usernames construction analysis matching first/last name of users: the first name ($w_1$) and/or last name ($w_2$) and digits ($d$).}
 \label{tab:nbu_partial_match}
\end{table}

The matching conditions are exclusive, following the order as presented in the
table (e.g., a username matching $w_1$ \emph{and} $w_2$ is counted in the
second row and not in the $w_1$ and $w_2$ rows separately). One of the most
remarkable results is that $70\%$ of the usernames contain at least one of the
two parts of the real name. In particular, $30\%$ of the collected usernames are
constructed by simply concatenating the first and last name without adding any
digit. We also observe that more than $18\%$ of the usernames are constructed
adding digits to the provided first and last names. This is most likely a
typical behavior of users, whose first chosen pseudonym (a variant of first
and last name) is not available and  that do add digits (e.g. birth year) to
be able to register into the service.


After this preliminary analysis we want to understand how $w_1$ and $w_2$ are
combined to build the exact username. In order to do that, we also consider the
first character of each word, namely $c_1$ and $c_2$ respectively.
Table~\ref{tab:nbu_exact_match} shows multiple ways to combine $w_1$, $w_2$,
$c_1$, $c_2$ and digits ($d$).  We provide the percentage of usernames observed
for each combination. The results show that more than 50\% of usernames match
exactly the patterns we tested.  One can observe that the most common way
usernames are generated from users' real names is by \textit{concatenating}
the first and last name, in that specific order (almost 14\%), or by adding
a dot between both names (~13\%).

\begin{table} [ht!]
  \centering
      \scriptsize
      \begin{tabular}{|c|c||c|c||c|c|}
        \hline
        Pattern & \% & Pattern & \% & Pattern & \% \\
        \hline
        \hline
      $w_1$$w_2$  &  13.99 &  $c_1$.$w_2$ &  0.44  &  $w_2$d   &  0.84\\
        \hline
      $w_2$$w_1$  &  1.46  &  $w_2$$c_1$  &  0.28  &  $w_1$$w_2$d &  6.04\\
        \hline
      $w_1$.$w_2$ &  12.95 &  $w_2$.$c_1$ &  0.08  &  $w_1$.$w_2$d   &  1.86\\
        \hline
      $w_2$.$w_1$ &  2.22  &  $c_2$$w_1$  &  0.09  &  $w_2$$w_1$d &  0.89\\
        \hline
      $w_1$ &  0.91  &  $w_1$$c_2$  &  0.44  &  $w_2$.$w_1$d   &  0.49\\
        \hline
      $w_2$ &  0.99  &  $w_1$.$c_2$ &  0.09  &  $c_1$$w_2$d &  2.58\\
        \hline
      $c_1$$w_2$  &  3.45  &  $w_1$d   &  2.71  &  $w_1$$c_2$d &  0.8\\
      \hline
      \end{tabular}
  \caption{Usernames exactly matching a pattern.}
 \label{tab:nbu_exact_match}
\end{table}



Again, because first-chosen usernames might be already in use, users typically
choose to (or are suggested to by the online service itself) add a number as a
postfix of their desired username. In particular, we observe that in most of
these cases, users add exactly respectively two or four numbers, in 40\% of
the cases and 20\% respectively. These ending digits suggest then either the
year of birth (full or simply the last two digits) or the birth date.


%

\begin{figure} [tb] \centering
\includegraphics[scale=0.65]{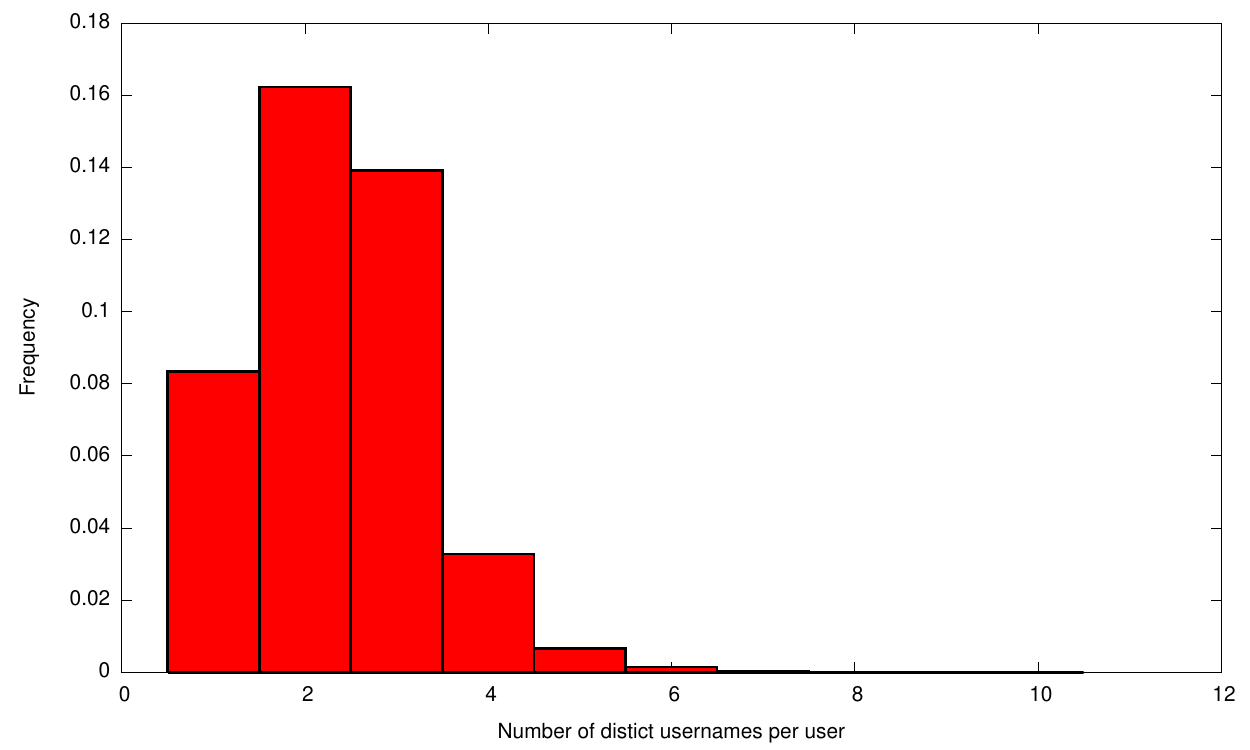}
\caption{Number of distinct usernames per user distribution}
\label{fig:n_usernames}
\end{figure}

Finally, figure \ref{fig:n_usernames} shows the distribution of the number of different usernames
the users in our dataset utilize. The graph shows that most users have two or three different usernames.
The mean number of usernames per user is $2.3$. 

\section{Discussion}
\label{sec:discussion}

Recently some governments and institutions are trying to pass laws and policies
to force users to tie their digital identities with their real ones. For
example, there is a current discussion in China and France \cite{france} 
on laws that would require users to use their real names when posting
comments on blogs and forums.  Similarly, the company Blizzard had started
an effort to tie real identities to the ones used to post comments on its
video games forums.

This work shows that it is clearly possible to tie digital identities together
and, most likely, to real identities in many cases only using ubiquitous
usernames. We also showed that, even though users are free to change their
usernames at will, they do not do it frequently and, when they do, it is in a
predictable way. Our technique might then be used as an additional tool when
investigating online crime. It is however also subject to abuse and could
result in breaches in privacy.  Advertisers could automatically build online
profiles of users with high accuracy and minimum effort, without the consent of
the users involved.

Spammers could gather information across the web to send
extremely targeted spam, which we dub {\em E-mail spam 2.0}.  For example, by
matching a Google profile and an eBay account one could send spam emails that
mention a recent sale. In fact, while eBay profiles do not show much personal
information (like real names) they do show recent transactions indexed by username.
This would enable very targeted and efficient phishing attacks. We argue
that these targeted attacks might have higher click rates for spammers thus
leading to smaller spam campaigns that would be much harder for spam
classifiers to recognize.

Finally, users could use our tool to assess how unique and linkable their
usernames are. They can thus take an informed decision on whether to change
their pseudonym for their online activity they wish to remain private.
Paradoxically, it would be difficult for an user who decides to prevent the
linking of her different usernames (particularly on OSNs), to choose usernames
that are unlinkable without loosing some of the
benefits of the various features of OSNs. 

In the light of our results, an analysis on the nature and anonymity of
usernames is needed. Historically usernames have been used to identify users in
small groups, one such example are Unix usernames. In groups of dozens or few hundreds of
people, usernames naturally tend to be not identifying and non unique \footnote{
We compared the mean entropy of the usernames on a shared server in our
lab with the ones gathered from Google and the difference is remarkable.}.
As the online communities grow in size, so does the entropy of the
usernames.  Nowadays users are forced to choose usernames that have to be
unique in online services that have hundreds of million of users. Naturally users had
to adapt and choose higher entropy usernames to be able to find usernames
that were not already assigned. This can allow for privacy breaches.


\subsection{Countermeasures}

\paragraph{On the user side}

Following this work users might change their username habits and use different
usernames on different web services.  We released our tool as a web application
that users can access to estimate how unique their username is and thus take
informed decision on the need to change their usernames when they deem
appropriate (\url{http://planete.inrialpes.fr/projects/how-unique-are-your-usernames}).

\paragraph{For web services}

There are two main features that make our technique possible and exploitable in real
case scenarios. First, web services and OSNs allow access to public accounts of their
users via their usernames. This can be used to easily check for existence of a given username
and to automatically gather information. Some web services like Twitter are built around
this particular feature.
Second, web services usually allow the user pages to be crawled automatically. While in some
cases this might be a necessary evil to allow search engines to access relevant content,
in many instances there is no legitimate use of this technique and indeed some OSNs explicitly 
forbid it in the terms of service agreements, e.g., Facebook.

While preventing automatic abuse of public content can be difficult in general, for example when the attacker has access to a large number of IPs, it
is possible to at least throttle access to those resources via CAPTCHAs \cite{captcha} or
similar techniques. For example, in our study we discovered that eBay presents
users with a CAPTCHA if too many requests are directed to their servers from the same IP.  

\section{Conclusion}

In this paper we introduced the problem of linking online profiles using
only usernames. Our technique has the advantage of being almost always
applicable since most web services do not keep usernames secret.
Two family of techniques were introduced. The first one estimates the
uniqueness of a username to link profiles that have the same username.  We
gather from language model theory and Markov-Chain techniques to estimate
uniqueness. 
Usernames gathered from multiple services have been shown to have a
high entropy and therefore might be easily traceable.


We extend this technique to cope with profiles that are linked but
have different usernames and tie our problem to the well known problem of
record linkage. All the methods we tried have high precision in linking
username couples that belong to the same users.

Ultimately we show a new class of profiling techniques that can be exploited
to link together and abuse the public information stored on online social
networks and web services in general.

\bibliographystyle{acm}
\bibliography{bibfile}

\appendix
\section*{Username uniqueness from a probabilistic point of view}
\label{app:unique}
We now focus on computing the probability that only {\em one} users
has chosen username $u$ in a population. We refer to this probability as $P_{uniq}(u)$. 

%

Intuitively $P_{uniq}(u)$ should increase with the decrease in likelihood of $P(u)$. However,
$P_{uniq}(u)$ also depends on the size of the population in which we are trying to estimate
uniqueness. For example, consider the case of first names. Even an uncommon first name
does not uniquely identify a person in a very large population, e.g. the US. However, it is very
likely to uniquely identify a person in a smaller population, like a classroom.

To achieve this goal we use the $P(u)$ to calculate the expected
number of users in the population that would likely choose username $u$.
Let us denote by $n(u)$ the expected number of users that choose string
$u$ as a username in a given population $W$.
The value of $n(u)$ is calculated based on $P(u)$ as: $$n(u)
= P(u)*W$$ \noindent where $W$ is the total number of users in the population.
In our case $W$ is an estimation of the number of users on the Internet: $1.93$ billions \footnote{
http://www.internetworldstats.com/stats.htm}.

In case we are sure there exists {\em at least one} user that selected the username $u$
(because $u$ is taken on some web service) then the computation of $n(u)$ changes slightly:
$$n(u) = P(u)*(W-1) + P(u|u) = P(u)*(W-1) + 1$$ \noindent where the addition of $1$ comes from the fact that
we are sure that there exists {\em at least} one user that choses $u$ and $W-1$ is there to account
for the person for which we are sure of.

Finally we can estimate the uniqueness of a username $u$ by simply considering
the probability that our user is unique in the reference set determined by $n(u)$, hence:
$$P_{uniq}(u) = \frac{1}{n(u)}$$

\end{document}